\documentclass[a4paper,11pt]{article}
\usepackage{pos}

\newcommand{\be}{\begin{equation}}
\newcommand{\ee}{\end{equation}}
\newcommand{\ba}{\begin{eqnarray}}
\newcommand{\ea}{\end{eqnarray}}
\newcommand{\nn}{\nonumber}

\newcommand{\rts}{\sqrt{s}}

\title{The $\phi p$ bound state in the unitary coupled-channel approximation}
\author*[a]{Bao-Xi Sun}
\author[a]{Ying-Ying Fan}
\author[a]{Qin-Qin Cao}

\affiliation[a]{School of Physics and Optoelectronic Engineering, Beijing University of Technology, \\
Beijing 100124, China}

\emailAdd{sunbx@bjut.edu.cn}

\abstract{The attractive interaction of the $\phi$ meson and the proton is reported by the ALICE Collaboration, and the corresponding scattering length $f_0$ is given as $Re(f_0)=0.85\pm0.34(stat)\pm0.14(syst)$ fm and $Im(f_0)=0.16\pm0.10(stat)\pm0.09(syst)$ fm. The fact that the real part is significant in contrast to the imaginary part indicates a dominating role of the elastic scattering, whereas the inelastic process is less important. In this work, such scattering processes are inspected on the basis of a unitary coupled-channel approximation inspired by the Bethe-Salpeter equation. The $\phi p$ scattering length is calculated and it is found that the experimental value of the $\phi p$ scattering length can be obtained only if the attractive interaction of the $\phi$ meson and the proton is taken into account. A significant outcome of such an attractive interaction is a two-pole structure in the scattering amplitude. One of the poles, located at $1969-i283$ MeV, might be a resonance state of $\phi N$, while the other pole, located at $1949-i3$ MeV, should be a bound state of $\phi N$. Both of these states do not have counterparts in the data of the Particle Data Group(PDG).}

\FullConference{The 21st International Conference on Hadron Spectroscopy and Structure (HADRON2025)\\
 27 - 31 March, 2025\\
Osaka University, Japan\\}

\begin{document}
\maketitle

\section{Introduction}

The strong interaction at low energies, due to its non-perturbative attribute, has led to an abundance of hadronic states. Understanding the nature of those hadrons poses a long-lasting challenge to physicists.
Remarkably, a recent result of the heavy-ion collision experiment at ALICE indicates an attractive interaction between the $\phi$ meson and the proton\cite{ALICE}.
However, the interaction of the vector meson with the baryon octet was studied in detail in \cite{ramos2010,Hosaka2011}.
These works suggested a vanishing elastic potential of the $\phi$ meson with the proton, and this conclusion is in contradiction to the recent ALICE result.
From the experiment, the interaction between the $\phi$ meson and the proton is shown to be attractive, as indicated by the corresponding scattering length given by the Lednicky-Lyuboshits fit. The real and imaginary parts of the scattering length read, respectively,
\begin{eqnarray}  
\label{eq:length}
{\rm Re}(f_0)&=&0.85\pm0.34\rm (stat)\pm0.14(syst)\,fm, \nn \\
{\rm Im}(f_0)&=&0.16\pm0.10\rm (stat)\pm0.09(syst)\,fm.
\end{eqnarray}
Compared to the error budget, the imaginary part of the scattering length is negligible.
This fact indicates that the elastic scattering plays a dominant role in the interaction between the $\phi$ meson and the proton, whereas the inelastic scattering is not important.
For the time being, the dynamical aspects of such a $\phi p$ system are still far from being understood and urgently require more studies.

The present work is dedicated to a further study on the dynamics of the $\phi p$ interacting system using an effective Lagrangian, in which 
both the vector meson and the baryon octet in the flavor $SU(3)$ space are incorporated.
Based on the Lagrangian, we will scrutinize the coupled-channel effects of such a system in more depths.
In particular, we will explore the impacts of an attractive $\phi p$ potential on the couple-channel dynamics.

\section{The hidden gauge formalism}
\label{sect:hidden}

The interaction of the vector meson with the baryon octet can be involved according to the hidden gauge formalism. The  Lagrangian of vector mesons can be written as
\be
\label{eq:vectorLag}
 L=-\frac{1}{4} \langle V_{\mu\nu} V^{\mu\nu} \rangle,
\ee
where the symbol $\langle \rangle$ represents the trace in the flavor $SU(3)$ space and $V_{\mu\nu}$ is the tensor of vector mesons,
\be
V_{\mu\nu} =\partial_\mu V_\nu-\partial_\nu V_\mu-ig \left[ V_\mu, V_\nu \right],
\ee
with $g=\frac{M_V}{2f}$ and the pion decay constant $f=93$MeV\cite{ramos2010}.

The Lagrangian of vector mesons in Eq.~(\ref{eq:vectorLag}) supplies an interacting vertex of three vector mesons in the interaction of the vector meson with the baryon octet, which comes from
\be
\label{eq:VVV}
L^{3V}_{III}=ig\langle \left( V^\mu \partial_\nu V_\mu -\partial_\nu V_\mu V^\mu \right) V^\nu \rangle,
\ee
with $V_\mu$ the $SU(3)$ matrix of vector mesons.

Similarly, the Lagrangian for the coupling of the vector meson to the baryon octet is constructed as
\be
\label{eq:BBV}
L_{BBV}=g\left(\langle \bar{B} \gamma_\mu  \left[V^\mu,B\right]\rangle+ \langle \bar{B} \gamma_\mu B \rangle \langle V^\mu \rangle   \right),
\ee
with $B$ the $SU(3)$ matrix of the baryon octet. 

With the interactions in Eqs.~(\ref{eq:VVV}) and (\ref{eq:BBV}), the $t-$ channel interaction of the vector meson and the baryon octet can be constructed, where a vector meson is exchanged between them.
Since the scattering process is studied on the energy region near the $\phi p$ threshold, the momentum of the exchanged vector meson can be neglected, and thus only the mass term is left in the propagator. Therefore, the potential of the vector meson with the baryon octet takes the form of
\be
\label{eq:202302041614}
V_{ij}=-C_{ij} \frac{1}{4f^2} \left(k^0+k^\prime{}^0 \right) \vec{\epsilon} \vec{\epsilon}^\prime,
\ee
where $k^0(\vec{\epsilon})$ and $k^\prime{}^0(\vec{\epsilon}^\prime)$ are energies(polarization vectors) of the incoming and outgoing vector mesons, respectively, and the $C_{ij}$ coefficient values can be obtained according to the Lagrangian in Eqs.~(\ref{eq:VVV}) and (\ref{eq:BBV}). 
The coefficients $C_{ij}$ for strangeness $S=0$ and isospin $I=1/2$ are listed in Table~\ref{s0i12}.
\begin{table}[htbp]
 \renewcommand{\arraystretch}{1.2}
\centering
\vspace{0.5cm}
\begin{tabular}{l|ccccc}
\hline\hline
 & $\rho N$ & $\omega N$ & $\phi N $  & $K^* \Lambda $ & $K^* \Sigma $\\
 \hline
$\rho N$ & 2 & 0 & 0 & $\frac{3}{2}$ & $-\frac{1}{2}$
  \\
$\omega N$ &  & 0 & 0 & $-\frac{3}{2}\frac{1}{\sqrt{3}}$ &
 $-\frac{3}{2}\frac{1}{\sqrt{3}}$ \\
$\phi N $ &  & & 0 & $-\frac{3}{2}
\left(-\sqrt{\frac{2}{3}}\right)$ &
 $-\frac{3}{2}
\left(-\sqrt{\frac{2}{3}}\right)$ \\
$K^* \Lambda $ &  & & & 0 & 0 \\
$K^* \Sigma $ &  & & & & 2  \\
\hline\hline
\end{tabular}
\caption{
 Coefficients $C_{ij}$ for strangeness $S=0$ and isospin $I=1/2$. 
}\label{s0i12}
\end{table}

\section{The unitary coupled-channel approximation}
\label{sect:BS}

A full scattering amplitude can be expressed in an integral representation according to the Bethe-Salpeter equation. By applying the on-shell condition to the potential $V$ involved, the equation can be reduced to a summation of bubble-loop series, 
\begin{eqnarray}
\label{eq:BS}
T&=&V+VGV+VGVGV+... \\ \nonumber
&=&\left[1-VG \right]^{-1} V,
\end{eqnarray}
with $G$ being the loop function of the intermediate vector meson and baryon.
According to this approximation, coupled-channel effects are well incorporated.
In the dimensional regularization scheme, the loop function $G$ takes the form of
\begin{eqnarray}
\label{eq:g-function}
  G_i(\rts) &=&  \frac{2 M_i}{(4 \pi)^2}
  \left\{
        a_i(\mu) + \log \frac{m_i^2}{\mu^2} +
        \frac{M_i^2 - m_i^2 + s}{2s} \log \frac{M_i^2}{m_i^2}
  \right.
  \\
     &+& \frac{Q_i(\rts)}{\rts}
    \left[
         \log \left(  s-(M_i^2-m_i^2) + 2 \rts Q_i(\rts) \right)
      +  \log \left(  s+(M_i^2-m_i^2) + 2 \rts Q_i(\rts) \right)
    \right.
  \nonumber
  \\
  & &
  \Biggl.
    \left.
      - \log \left( -s+(M_i^2-m_i^2) + 2 \rts Q_i(\rts) \right)
      - \log \left( -s-(M_i^2-m_i^2) + 2 \rts Q_i(\rts) \right)
    \right]
  \Biggr\},
  \nonumber
\end{eqnarray}
where the regularization scale is $\mu=630$ MeV and the subtraction constant will be determined with the $\phi N$ scattering length. In the calculation, the influence of the decay widths of the $\rho$ and $K^*$ mesons is taken into account.

\section{The Yukawa-type potential of the $\phi$ meson with the proton }
\label{sect:Yukawa}

The Yukawa-type potential of the $\phi$ meson with the proton takes the form of
\be
\label{eq:Yukawa}
V(r)=-A \frac{\exp(-\alpha r)}{r},
\ee
with $A=0.021\pm0.009\pm0.006(syst)$ and $\alpha=65.9\pm38.0(stat)\pm17.5(syst)$ MeV\cite{ALICE}.
Therefore, the $\phi p$ potential in the momentum space can be obtained with a Fourier transformation of Eq.~(\ref{eq:Yukawa}), which is written as
\be
\label{eq:Yukawamomen}
V(\vec{q})=\frac{-g^2_{\phi N}}{\vec{q}^2+\alpha^2},
\ee
with $\frac{g^2_{\phi N}}{4 \pi}=A$.
Apparently, the potential in Eq.~(\ref{eq:Yukawamomen}) is a non-relativistic form assuming that the three-momentum of the proton($\phi$-meson) is far less than the proton($\phi$-meson) mass. In this case, the zero component of the momentum transfer in the $t-$channel interaction of $\phi p$ tends to zero in the nonrelativistic approximation\cite{Sun:2022cxf}. It should be noted that the $\phi p$ coupling in this work $g_{\phi N}^2/4\pi=A$ is different from that of~\cite{ALICE}, where $g_{\phi N}^2=A$.

When the Bethe-Salpeter equation is solved, the Yukawa-type potential in Eq.~(\ref{eq:Yukawamomen}) is multiplied by a mass of the $\phi$ meson.
and the three-momentum transfer $\vec{q}$ vanishes at the threshold of $\phi p$. In reality, we calculated the effect caused by the three-momentum transfer $\vec{q}$ in the $\phi p$ potential, and no significant impact on the final results is observed.
Therefore,
\be
\label{eq:Yukawamomenzero}
V(\vec{0})=\frac{-g^2_{\phi N} m_\phi}{\alpha^2}.
\ee
In addition to the kernel in Eq.~(\ref{eq:202302041614}), the potential in Eq.~(\ref{eq:Yukawamomenzero}) is also taken into account when the Bethe-Salpeter equation is solved.
In the calculation of the present work, we set $A=0.021$ and $\alpha=65.9$MeV. 

\section{The $\phi N$ bound state}
\label{sect:bound}

The scattering amplitude $T$ of the $\phi N\rightarrow \phi N$ reaction is evaluated by solving the Bethe-Salpeter equation in the unitary coupled-channel approximation, and then the $\phi N$ scattering length can be obtained according to the $\phi N$ elastic scattering amplitude at the threshold, i.e.,
\be
a_{\phi N}=\frac{M_N}{8\pi \sqrt{s}}T_{\phi N\rightarrow \phi N}(\sqrt{s}=m_\phi+M_N).
\ee

To obtain the scattering length of $\phi N$ consistent with the experimental value provided by the ALICE Collaboration, we include the attractive $\phi N$ potential as given by Eq.~(\ref{eq:Yukawamomenzero}) when the Bethe-Salpeter equation is solved.
Adjusting the subtraction constant values, the $\phi N$ scattering length is obtained and is consistent with the experimental value in the uncertainty range. The subtraction constant values and the corresponding $\phi N$ scattering length are listed in Table~\ref{table:subS0I12}.

\begin{table}[htbp]
\centering
\begin{tabular}{ccccc|cc}
\hline\hline
       $a_{\rho N}$&$a_{\omega N}$&$a_{\phi N}$&$a_{K^* \Lambda}$&$a_{K^* \Lambda}$ & $f_0$ (MeV) \\
\hline
        -2.0  &  -2.0  &  -2.4  &  -1.9  &  -1.8  &  0.86+i0.19  \\
\hline \hline
\end{tabular}
\caption{The subtraction constants and the real and imaginary parts of the corresponding $\phi N$ scattering length $f_0$ obtained in the unitary coupled-channel approximation when the $\phi p$ attractive interaction is taken into account, where the regularization scale $\mu=630$MeV.} \label{table:subS0I12}
\end{table}

With the subtraction constants listed in Table~\ref{table:subS0I12}, the scattering amplitudes $|T_{ii}|^2$ of the vector meson and the baryon octet are calculated in the unitary coupled-channel approximation, and a peak around 1950 MeV is visible in the $\phi N$, $K^* \Lambda$ and $K^* \Sigma$ channels. Since it is below the $\phi N$ threshold, it could be associated with a $\phi N$ bound state.
Moreover, another pole of $|T_{ii}|^2$ is detected around 1970 MeV in the $K^* \Lambda$ and $K^* \Xi$ channels. Since it is higher than the threshold of $\phi N$, it can be regarded as a resonance state of $\phi N$.
Both of these two states are degenerate in $J^P=1/2^-$ and $J^P=3/2^-$, and do not have counterparts in the PDG data.

The pole positions and the corresponding couplings to different channels are shown in Table~\ref{table:coupling}.
Since the inelastic scattering amplitudes are far lower than the elastic ones, all couplings are calculated according to elastic processes.
It can be seen that the state at $1949-i3$ MeV is strongly coupled to the $\phi N$, $K^* \Lambda$ and $K^* \Sigma$ channels, while the state at $1969-i283$ MeV is mainly coupled to the $K^* \Lambda$ and $K^* \Sigma$ channels.

\begin{table}[htbp]
\centering
\begin{tabular}{c|c|c|c|c|c}
\hline\hline
 Pole positions (MeV)  &  $\rho N$          &  $\omega N$   &   $\phi N$          & $K^* \Lambda$ & $K^* \Sigma$ \\
\hline
$1949-i3$   & $0.0+i0.0$  &  $0.1+i0.0$ &  $2.1+i0.1$ & $1.6+i0.3$ & $1.8+i0.0$  \\
$1969-i283$ & $0.1+i0.1$  &  $0.0+i0.2$ &  $0.1-i0.1$  & $0.3-i0.4$   & $0.2-i0.0$  \\
\hline \hline
\end{tabular}
\caption{Pole positions and coupling constants to various channels of strangeness $S=0$ and isospin $I=1/2$.}
\label{table:coupling}
\end{table}

The bound state of $\phi p$ has been reconfirmed in~\cite{Chizzali:2022pjd}, where the experimental correlation function of $\phi p$ has been fitted with the constraint of the lattice calculation results. Eventually, the vector meson-baryon octet interaction without the attractive $\phi p$ potential is reconsidered in the unitary coupled-channel approximation, and the $\phi p$ correlation function is calculated to fit the experimental data\cite{Feijoo:2024bvn}. However, no $\phi N$ bound state is produced because the $\phi p$ scattering amplitude is spin-degenerate. In ~\cite{Abreu:2024qqo}, the spin $1/2$ and spin $3/2$ correlation functions of $\phi p$ are calculated independently, and the spin-averaged combination fits well with the experimental data. Moreover, a bound state of $\phi N$ is found in the case of spin $3/2$.  
Apparently, a new method to determine the subtraction constants in the loop function is proposed when the Bethe-Salpeter equation is solved. In addition, the relation of the study of hadronic resonance states with high energy heavy-ion collision experiments is being considered\cite{ExHIC:2017smd}. 

\section{Summary}
\label{sect:summary}

In the present work, we investigate the implications of an attractive interaction of the $\phi$ meson with the proton, as reported by the ALICE Collaboration.
An effective description is obtained by introducing a Yukawa-type potential on top of the vector meson and baryon octet interaction given by the effective Lagrangian.
We deliberate on the impacts of such a potential on the coupled-channel effects of the vector meson and baryon octet scattering system,
and a unitary approximation has been used to implement the coupled-channel effects. This approximation involves a reduced summation of the Bethe-Salpeter equation, and we arrive at a $\phi N$ scattering length consistent with the experiment data.
Special attention has been paid to the coupled channel effects of the vector meson and baryon octet scattering system with strangeness $S=0$ and isospin $I=1/2$, and two poles of the scattering amplitudes have been found in the complex energy plane. 
One pole is located at $1969-i283$ MeV and might be a resonance state of $\phi N$. 
In addition, we predict a possible bound state of $\phi N$, which has not been found in the PDG data. This state leads to another pole structure at $1949-i3$MeV in the scattering amplitude.

\end{document}